\documentclass[a4paper]{jpconf}
\usepackage{graphicx}
\usepackage{amsmath}
\newcommand{\lagr}{\mathcal{L}}

\begin{document}
\title{Astroparticle Physics Tests of Lorentz Invariance Violation}

\author{R G Lang and V de Souza}

\address{Instituto de F\'isica de S\~ao Carlos, Universidade de S\~ao Paulo. Avenida Trabalhador S\~ao-Carlense, 400, S\~ao Carlos, SP, Brazil}

\ead{rodrigo.lang@usp.br}

\begin{abstract}
Testing Lorentz invariance is essential as it is one of the pillars of modern physics. Moreover, its violation is foreseen in several popular Quantum Gravity models. Several authors study the effects of Lorentz invariance violation (LIV) in the propagation of ultra-high energy cosmic rays. These particles are the most energetic events ever detected and therefore represent a promising framework to test LIV. In this work we present an analytic calculation of the in-elasticity for any $a+b \rightarrow c+d$ interaction using first order perturbation in the dispersion relation that violates Lorentz invariance. The inelasticity can be calculated by solving a third-order polynomial equation containing: a) the kinematics of the interaction, b) the LIV term for each particle and c) the geometry of the interaction. We use the inelasticity we calculate to investigate the proton propagation in the intergalactic media. The photopion production of the proton interaction with the CMB is taken into account using the inelasticity and the attenuation length in different LIV scenarios. We show how the allowed phase space for the photopion production changes when LIV is considered for the interaction. The calculations presented here are going to be extended in order to calculated the  modified ultra-high energy cosmic rays spectrum and compare it to the data.
\end{abstract}

\section{Introduction}
Relativity is one of the most important theories of modern physics and one of its pillars is the Lorentz invariance, which is proposed to be a fundamental symmetry in Quantum Field Theory. The possibility of Lorentz invariance violation (LIV), however, has been investigated by several quantum gravity (QG) models and is incorporated in high energy models of space-time structure~\cite{Mattingly2005}. Testing LIV is, therefore, very relevant in order to investigate new proposals and also to set the range of validity of relativity.

In this work, we study the possibility of testing LIV with the propagation of ultra-high energy cosmic rays (UHECR). We use a generic first order perturbative term in the free particle Lagrangian~\cite{ColemanGlashow}, which results in a new momentum dependent term in the dispersion relation. This correction is expected to be suppressed by the Planck Scale. Thus, astrophysics is a good framework for testing it, since UHECR have been detected by the Pierre Auger Observatory~\cite{AUGER} with energies three orders of magnitude larger than the events detected in particle accelerator.

Ultra-high energy cosmic rays interact with the low energy photon background as they propagate through the Universe. Protons interact mainly with the Cosmic Microwave Background (CMB) via pion and pair production~\cite{Stecker1968,Blumenthal1970}, heavier nuclei also interact with the Extragalactic Background Light (EBL), undergoing photo-disintegration~\cite{Stecker1999} and energetic photons can produce an electron-positron pair when interacting with the CMB and the Radio Background (RB)~\cite{DeAngelis2013}. The main structures in the UHECR spectrum could be caused by these interactions. The dip model proposes that in a pure proton spectrum, the ankle at $10^{18.7}$ eV appears naturally due to pair production~\cite{Berezinsky2009}. On the other hand, one of the proposed models to explain the suppression at the highest energies could be due to pion production and photo-disintegration, which is called the GZK suppression~\cite{Greisen,ZatsepinKuzmin}. If LIV is considered, the energy losses of the UHECR in the interactions with the CMB are changed and, consequently, the energy spectrum structures too.  For this reason, the UHECR spectrum, currently measured by the Pierre Auger Observatory~\cite{AUGERSpectrum} and by the Telescope Array experiment~\cite{SpectrumTA}, can be a reliable source of information to impose limits in the LIV.

This paper is organized as follows: in Section 2 we present the LIV framework used. In Section 3 we discuss the modifications in the kinematics of interactions of propagating UHECR. In Section 4 we show the results in the photopion production. And, finally, in Section 5 we conclude the paper and discuss other important tests that could be studied.

\section{Lorentz Invariance Violation}

The Coleman and Glashow formalism, proposed in~\cite{ColemanGlashow}, is a formulation which has been used in several works to study cosmic rays propagation with LIV~\cite{Jacobson2002,Stecker2005,Maccione2009,Scully2009}. It proposes a perturbation to the free particle Lagragian:

\begin{equation}
\centering
\lagr = \partial_{\mu} \psi^{*} Z \partial^{\mu} \psi - \psi^{*} M^{2} \psi
\end{equation}

\begin{equation}
\centering
\lagr \rightarrow \lagr + \partial_{i} \psi \delta_{a} \partial^{i} \psi
\end{equation}

This new perturbative term must be calculated for each particles and it leads to a new dispersion relation given by:

\begin{equation}
\centering
E_{a}^{2} = p_{a}^{2} c^{2} + m_{a}^{2} c^{4} + \delta_{a} p_{a}^{2} c^{2}
\end{equation}

which can be rewritten as:

\begin{equation}
\centering
E_{a}^{2} = p_{a}^{2} c^{2} (1+\delta_{a}) + m_{a}^{2} c^{4}
\end{equation}

It is then defined $c_{a} = c \sqrt{1+\delta_{a}}$:

\begin{equation}
\centering
E_{a}^{2} = p_{a}^{2} c_{a}^{2} + m_{a}^{2} \frac{c_{a}^{4}}{(1+\delta_{a})^{2}} \approx p_{a}^{2} c_{a}^{2} + m_{a}^{2} c_{a}^{4}
\end{equation}

therefore, $c_{a}$ is the maximum attainable velocity of a particle or its ``own speed of light''.

The main effect of this perturbation is a shift in the center of mass system (cms) energy of a particle\footnote{We were expressly showing $c$ in the dispersion relation in order to compare with $c_{a}$, but from now on we will use natural units where $c = 1$.}:

\begin{equation}
\label{eq:DispersionRelation}
\centering
\sqrt{s_{a}} = \sqrt{E_{a}^{2} - p_{a}^{2}} = \sqrt{m_{a}^{2} + \delta_{a} p^{2}_{a}}
\end{equation}

The cms energy is crucial for the calculation of the energy threshold of interactions and, consequently, a shift in this quantity results in a shift in the threshold of such interactions.

\section{Kinematics of Propagating UHECR}

In order to calculate the effects of LIV in the energy spectrum of UHECR, it is necessary to calculate the energy losses of propagating cosmic rays considering LIV. In this work we present the calculation for any generic $a + b \rightarrow c + d$ interaction based on the calculations performed by Scully and Stecker \cite{Scully2009}.

Scully and Stecker assume the effects of LIV only in the inelasticity, leaving the cross section unchanged. The same assumptions are used in this work. Their calculations for the inelasticity, however, are performed in the nucleus reference frame (NRF) and, then, a boost to the laboratory frame\footnote{The laboratory frame (LF) is that in which the CMB is isotropic and is the frame in which the energy spectra of cosmic rays are usually presented.} is necessary. Consequently, another assumption, that the boost is not modified by LIV is needed. The following calculations are already performed in the LF and, therefore, this assumption is avoided.

The inelasticity is obtained by imposing that the total rest energy, $\sqrt{s} = \sqrt{E^{2}_{tot} - p^{2}_{tot}}$ is conserved, i.e., $s_{i} = s_{f}$. We can, then, obtain $s_{i}$ and $s_{f}$ in terms of the properties of the particles $a$, $b$, $c$ and $d$:

\begin{equation}
\centering
\begin{cases}
s_{i} = \left(E_{a}+E_{b}\right)^{2} - \left(\vec{p}_{a}+\vec{p}_{b}\right)^{2} = s_{a} + s_{b} + 2 E_{a} E_{b} - 2 p_{a} p_{b} \cos{\theta_{i}} \\
s_{f} = \left(E_{c}+E_{d}\right)^{2} - \left(\vec{p}_{c}+\vec{p}_{d}\right)^{2} = s_{c} + s_{d} + 2 E_{c} E_{d} - 2 p_{c} p_{d} \cos{\theta_{f}}
\end{cases}
\end{equation}

We then rewrite the momentum as a function of the energy and the mass and use that $E^{2} >> s$:

\begin{equation}
\centering
s_{i} = s_{a} + s_{b} + 2 E_{a} E_{b} \left(1 - \cos{\theta_{i}} + \cos{\theta_{i}} \left(\frac{s_{a}}{2 E^{2}_{a}} + \frac{s_{b}}{2 E^{2}_{b}}\right) + \mathcal{O}(2) \right)
\end{equation}

\begin{equation}
\centering
s_{f} = s_{c} + s_{d} + 2 E_{c} E_{d} \left(1 - \cos{\theta_{f}} + \cos{\theta_{f}} \left(\frac{s_{c}}{2 E^{2}_{c}} + \frac{s_{d}}{2 E^{2}_{d}}\right) + \mathcal{O}(2) \right)
\end{equation}

In the regime where $a$ is a relativistic particle interacting with a low energy background and generating two relativistic particles, which is valid for all the interactions of interest\footnote{The pair production produces three particles: an electron, a positron and a photon. The electron-positron pair, however, can be approximated as a single particle with double the mass for the calculations.}, the final angle can be approximated as $\cos{\theta_{f}} \approx 1$, therefore:

\begin{equation}
\centering
s_{f} \approx s_{c} + s_{d} + 2 E_{c} E_{d} \left(\cos{\theta_{f}} \left(\frac{s_{c}}{2 E^{2}_{c}} + \frac{s_{d}}{2 E^{2}_{d}}\right)\right)
\end{equation}

Finally, we want to look at the inelasticity of the interaction that is defined as the fraction of energy lost by the particle $a$ due to the creation of $c$. In this regime, we can write:

\begin{equation}
\centering
K = \frac{E_{c}}{E_{a}} \implies E_{c} = K E_{a}
\end{equation}

\begin{equation}
\centering
K = 1 - \frac{E_{d}}{E_{a}} \implies E_{d} = (1-K) E_{a}
\end{equation}

Substituting $E_{c}$ and $E_{d}$ in $s_{i} = s_{f}$ and writing the cms energy of each particle using Eq.~\ref{eq:DispersionRelation}, we have:

\begin{equation}
\centering
\begin{array}{c}
s_{a} + s_{b} + 2 E_{a} E_{b} \left(1 - \cos{\theta_{i}} + \cos{\theta_{i}} \left(\frac{s_{a}}{2 E_{a}} + \frac{s_{b}}{2 E_{b}}\right)\right) \\ = s_{c} (K) + s_{d} (K) + K (1-K) \left(\cos{\theta_{f}} \left(\frac{s_{c} (K)}{K} + \frac{s_{d} (K)}{(1-K)}\right) \right)
\end{array}
\end{equation}

It is highly desirable to write this expression as a polynomial, for it is easier to solve. Therefore, we first define:

\begin{equation}
\centering
A := s_{i} - m_{c}^{2} - m_{d}^{2}
\end{equation}

and, then, multiply the expression\footnote{This is only possible as we know that $K \neq 0$ and $K \neq 1$. In the first situation there is no interaction and, in the second, only a pion is created, which is physically forbidden.} by $K(1-K)$:

\begin{equation}
\centering
\begin{split}
-m_{c}^{2} + K \left(A+2m_{c}^{2} - \delta_{d} E_{a}^{2}\right) + K^{2} \left(-A-m_{c}^{2}-m_{d}^{2} - \delta_{c} E_{a}^{2}\right) + K^{3} \left(\delta_{c} E_{a}^{2} \right) + \\
- \delta_{d} E_{a}^{2} \sum_{i=0}^{3} K^{i+1}  \binom{3}{i} \left(-1\right)^{i} - \delta_{d} E_{a}^{2} \sum_{i=0}^{2} K^{i} \binom{2}{i} \left(-1\right)^{i+2} = 0
\end{split}
\end{equation}

This third-order polynomial contains all the kinematics, LIV and geometry information and is easily solvable. Nevertheless, only roots in the regions $(0,1)$ are physical solutions for the problem.

Lastly, we use the modified inelasticity to calculate the attenuation length, $\ell$, which is defined as the distance the cosmic ray can travel before losing $1/e$ of its energy:

\begin{equation}
\centering
\frac{1}{\ell (E)} = \frac{1}{E} \frac{dE}{dx} = \int_{0}^{\infty} d\epsilon n(\epsilon) \sigma(\epsilon) \int_{-1}^{1} dcos \left(\theta_{i}\right) K(\epsilon,E,\theta_{i})
\end{equation}

where $n(\epsilon)$ is the background density, $\sigma(\epsilon)$ is the cross section and the threshold appears naturally in the inelasticity, as $K = 0$ for $\epsilon < \epsilon_{th}$.

With this information it is possible to calculate the propagation for any cosmic ray including LIV effects. In the next section we show the results for protons producing pions ($p + \gamma \rightarrow p + \pi$) as an example.

\section{Photopion Production}

The photopion production with LIV has been widely studied \cite{AlfaroPalma,Stecker2005,Scully2009,Xiao-Jun2009,Bietenholz,Saveliev,Cowsik2012,Boncioli2015,Aloisio2000} as this interaction modulates the shape of the spectrum in the highest energies, where the LIV effects could be strong enough to be detected. We have calculated the inelasticity and the attenuation length for this interaction using different values of $\delta_{\pi}$ and $\delta_{p}$ as an example for the calculations presented in the last section.

\begin{figure}[h]
\centering
\begin{minipage}{11.5pc}
\includegraphics[width=11.5pc]{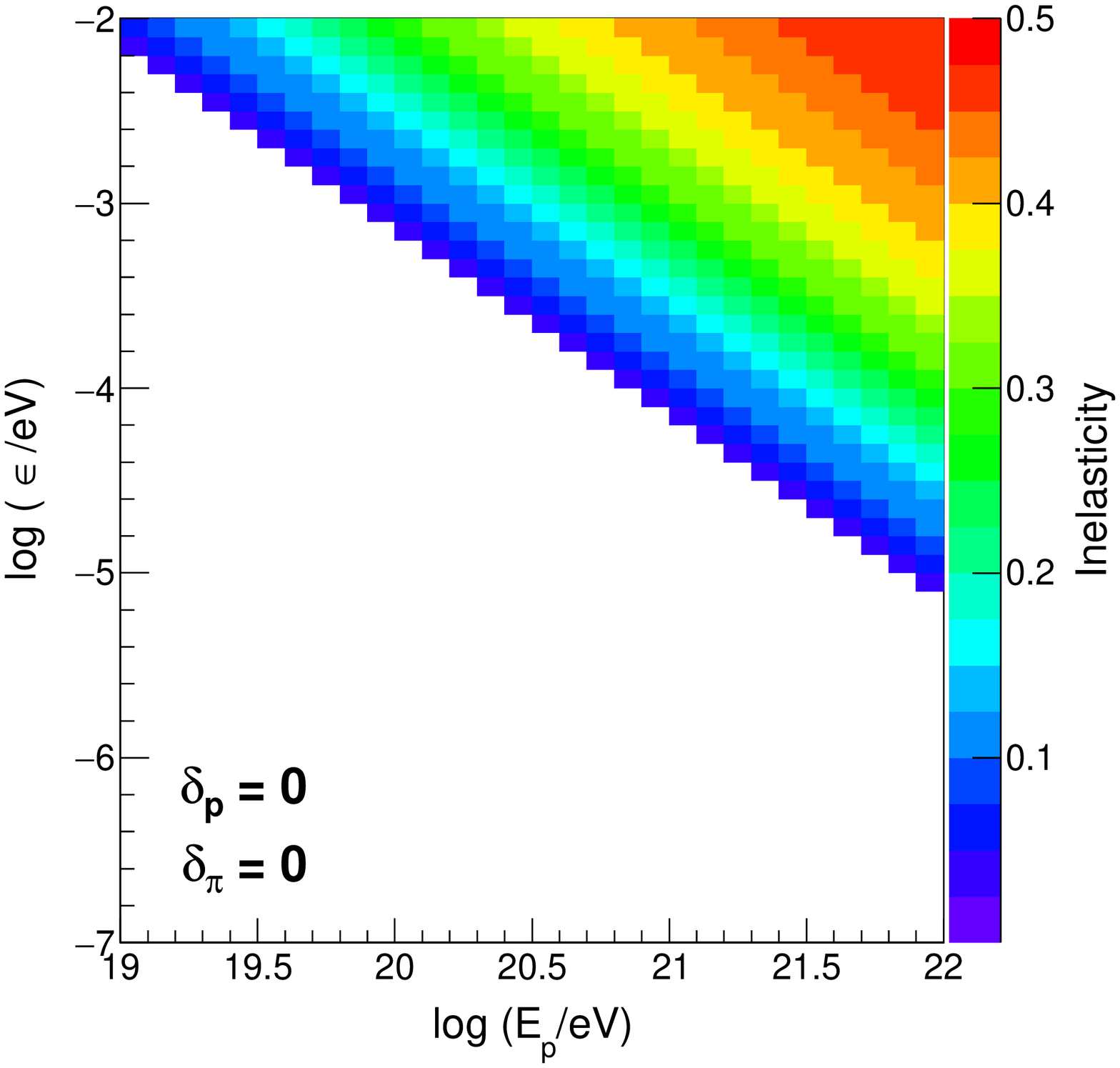}
\caption{\label{fig:Inelasticity0_0}Inelasticity for the photopion production in a scenario without LIV.}
\end{minipage}\hspace{1.5pc}%
\begin{minipage}{11.5pc}
\includegraphics[width=11.5pc]{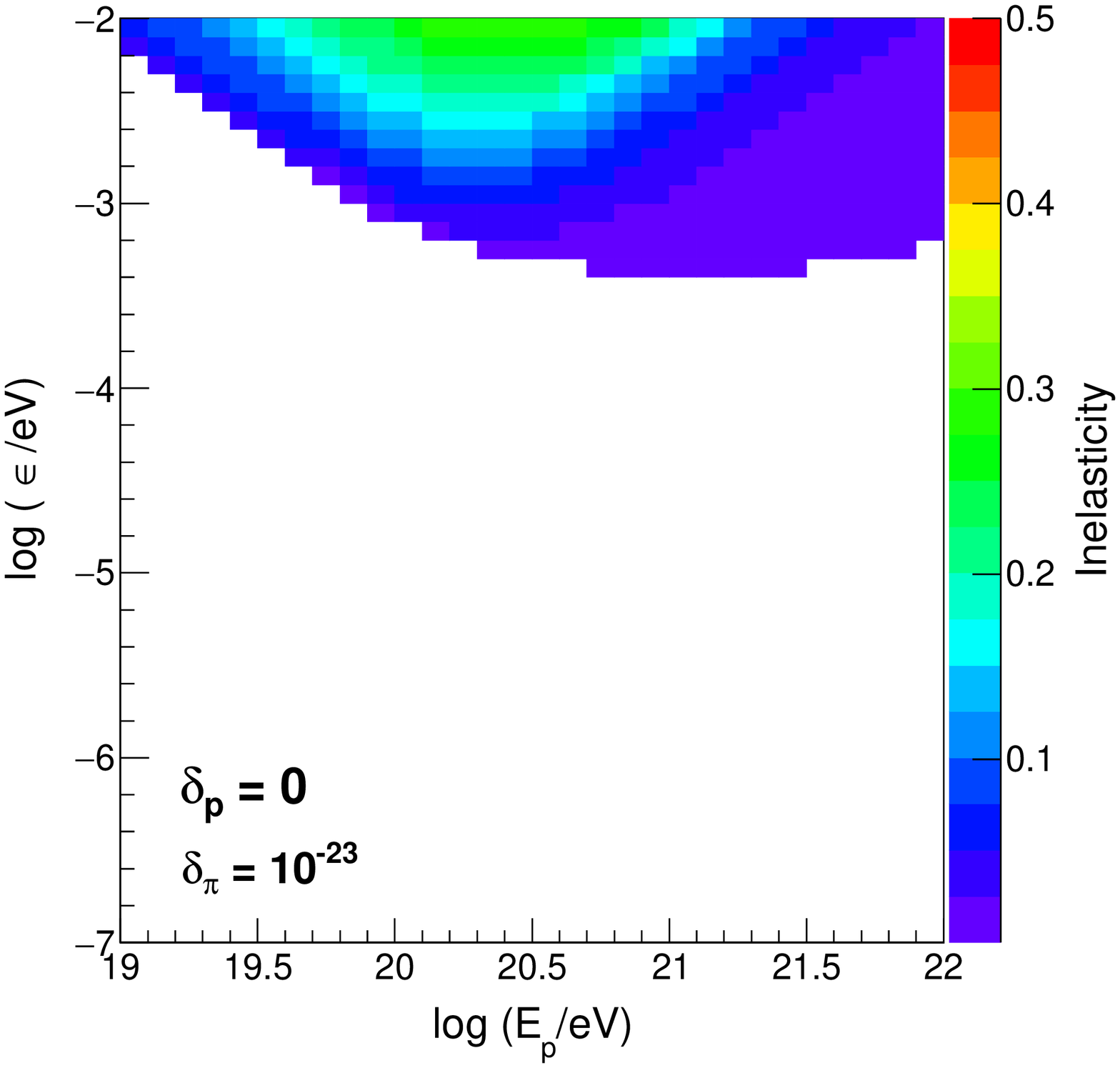}
\caption{\label{fig:Inelasticity10_0}Inelasticity for the photopion production in a scenario with LIV for the pion.}
\end{minipage} \hspace{1.5pc}%
\begin{minipage}{11.5pc}
\includegraphics[width=11.5pc]{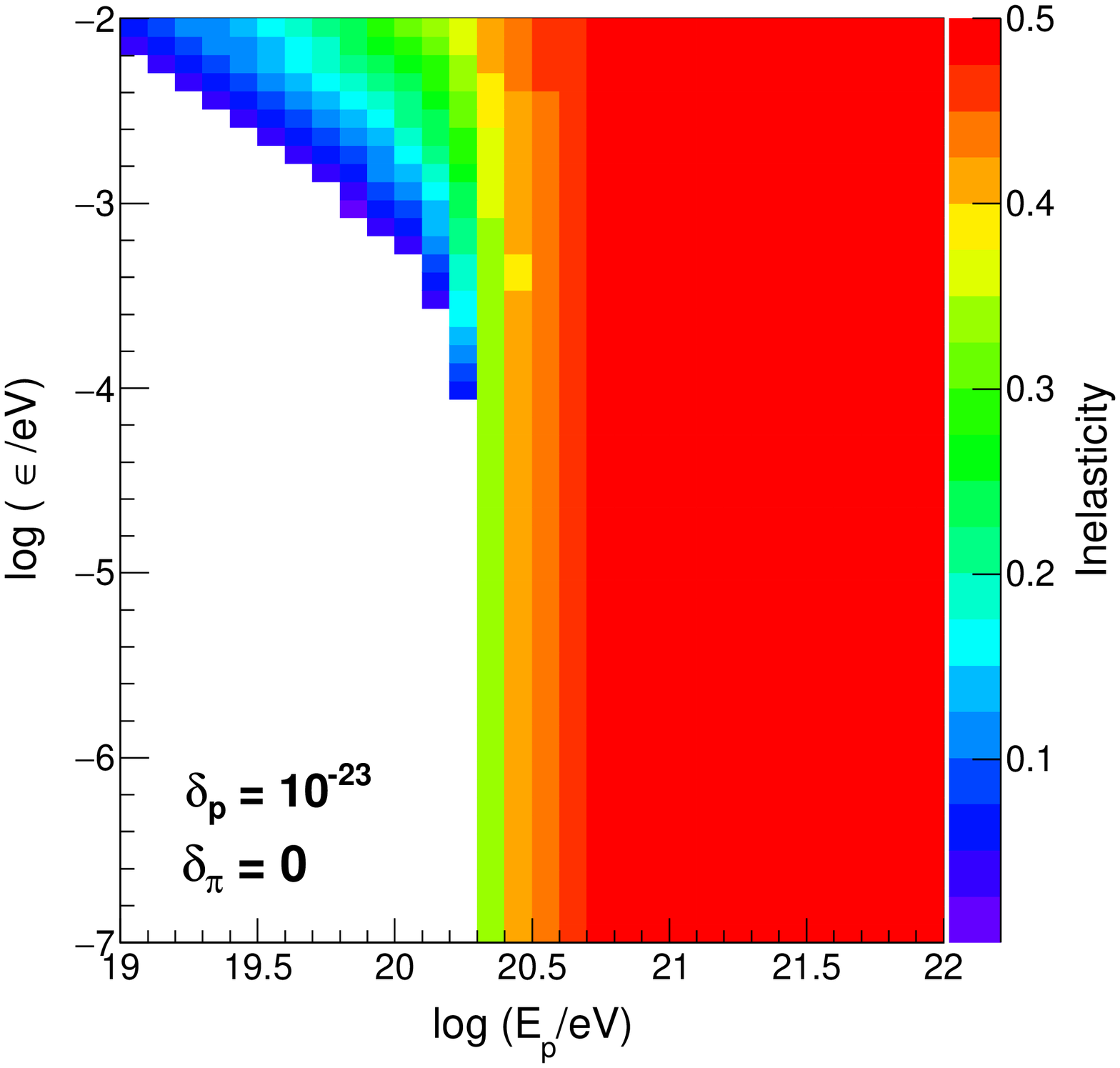}
\caption{\label{fig:Inelasticity0_10}Inelasticity for the photopion production in a scenario with LIV for the proton.}
\end{minipage}
\end{figure}

Figs. \ref{fig:Inelasticity0_0}-\ref{fig:Inelasticity0_10} show the inelasticity for the interaction as a function of the initial proton energy and the background photon energy, both in the LF. These figures also show the threshold energy for the background photon, which is the first energy with non-zero inelasticity, and the phase space of the interaction.

Turning on LIV for the pion reduces the phase space for the interaction as the threshold for the interaction becomes larger when compared to the LI case. This effect gets stronger as the initial proton energy gets larger. The opposite happens for the proton, the threshold is reduced, resulting in a larger phase space.

\begin{figure}[h]
\centering
\begin{minipage}{16pc}
\includegraphics[width=16pc]{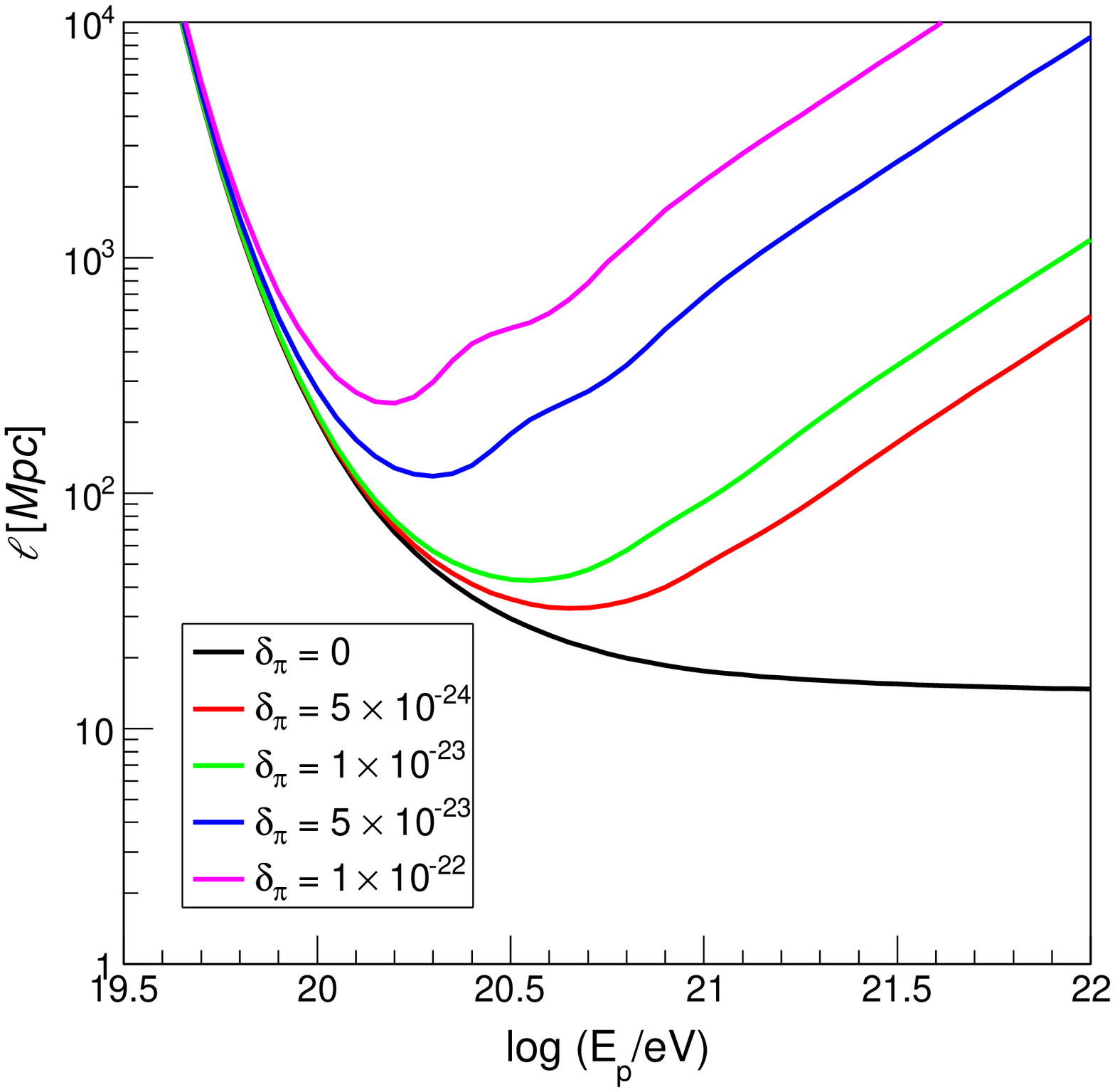}
\caption{\label{fig:AttLengthPion}Attenuation length for the photopion production with LIV for the pion. Each line show the result for a different LIV term.}
\end{minipage}\hspace{2pc}%
\begin{minipage}{16pc}
\includegraphics[width=16pc]{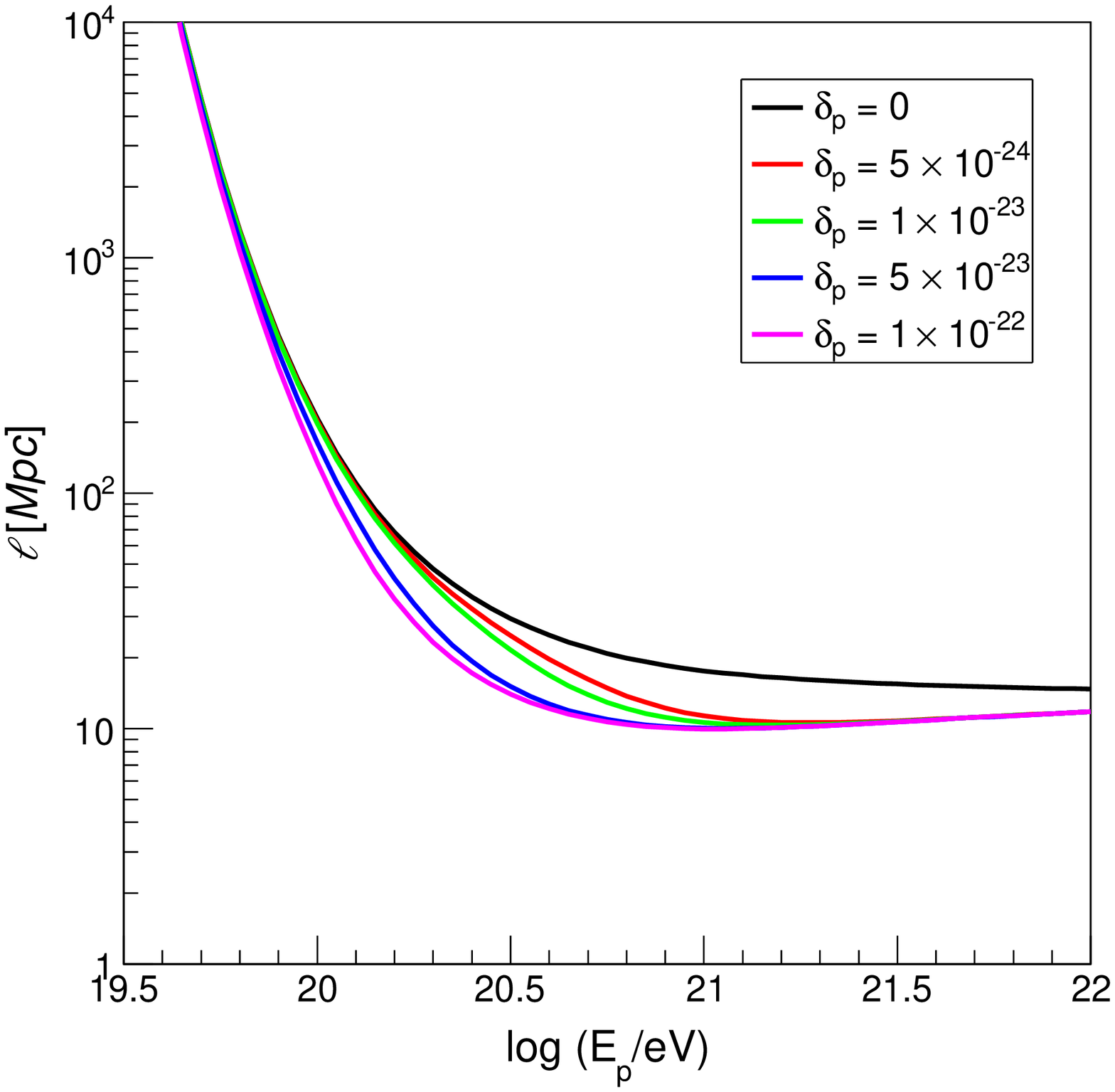}
\caption{\label{fig:AttLengthProton}Attenuation length for the photopion production with LIV for the proton. Each line show the result for a different LIV term.}\end{minipage}
\end{figure}

Fig. \ref{fig:AttLengthPion} and Fig. \ref{fig:AttLengthProton} show the attenuation length, i.e., the distance the proton can travel before losing $1/e$ of its energy, for different LIV terms. The changes in the phase space have direct effects in this. For the pion, the attenuation length gets larger and, therefore, the cosmic ray travels further before interacting. For the proton, on the other hand, it gets smaller, which means that the cosmic ray will interact more often.

The LIV term in the dispersion relation can be treated as an effective mass that gets larger with the energy. For the pion, a larger mass would mean more energy needed to create it, consequently shifting up the threshold. A shift in the proton mass, on the other hand, would change the energy both before and after the interaction. But, as the initial proton is more energetic, the shift before the interaction is larger and therefore the energy of the background photon necessary for creating the pion is smaller.

Those changes have direct influence in the UHECR spectrum as it could be recovered or suppressed in the highest energies.

\section{Conclusions}

The propagation of ultra energetic cosmic rays is a reliable source for testing Lorentz Invariance Violation as it leaves footprints in the energy spectrum measured with high statistics by different experiments.

In this work we have presented a generic analytic calculation for the inelasticity in the LF of any $a + b \rightarrow c + d$ interaction. The inelasticity is directly related to the threshold of these interactions and is used to obtain their energy losses. We have shown, as an example, the results of the calculations for the photopion production. The main effect is a change in the phase space of the interaction, which can be larger or smaller, depending on the LIV terms.

The calculations here proposed can be used in several LIV studies. Limits for the LIV terms, for example, could be imposed by performing these calculations for all the interactions present in the propagation and obtaining the spectrum using propagation codes such as CRPropa~\cite{CRPropa3} and SimProp~\cite{SimProp}.

\ack
We would like to thank the S\~ao Paulo Research Foundation (FAPESP) for the financial support through Grant No. 2014/26816-0 and 2010/07359-6 and Conacyt for the support through the Mesoamerican Center for Theoretical Physics, provinding a scholarship to Rodrigo Guedes Lang in order to participate in the school and present this work. We would also like to thank Humberto Martinez for the help provided to fully understand and double check the calculations.

\section*{References}
\bibliography{bibtex.bib}

\end{document}